\documentclass[12pt,preprint]{aastex}
\newcommand{\ks}{km s$^{-1}$~}
\newcommand{\kms}{km s$^{-1}$~}
\newcommand{\HI}{H{\sc i}}

\newcommand{\g}{$^\circ$}

\newcommand{\cmm}{cm$^{-3}$}
\newcommand{\vlsr}{V$_{\rm LSR}$}
\slugcomment{Accepted, Astronomical J.}
\begin{document}

\title{Investigation of the large-scale neutral hydrogen near the supernova 
remnant W28}

\author{P. F. Vel\'azquez\altaffilmark{1}}
\affil{Instituto de Ciencias Nucleares, Universidad Nacional Aut\'onoma de
M\'exico, \\ 
Apdo. Postal 70-543, C.P.: 04510, M\'exico D.F., M\'exico\\ 
e-mail: pablo@nuclecu.unam.mx}

\author{G.M. Dubner\altaffilmark{2}}
\affil{Instituto de Astronom\'\i a y F\'\i sica del Espacio
(CONICET, UBA),\\ 
C.C.67, 1428 Buenos Aires, Argentina.\\
e-mail: gdubner@iafe.uba.ar}

\author{W. M. Goss}
\affil{National  Radio  Astronomy  Observatory,\\
P.O.Box 0,Socorro, New Mexico 87801, USA.\\
e-mail: mgoss@nrao.edu}

\and

\author{A. J. Green}
\affil{School of Physics, University of Sydney,\\
NSW 2006, Australia.\\
e-mail: agreen@physics.usyd.edu.au}

\altaffiltext{1}{Fellow of CONICET, Argentina}
\altaffiltext{2}{Member of the Carrera del Investigador Cient\'\i fico,
CONICET, Argentina}

\begin{abstract}
  The distribution and kinematics of neutral hydrogen have been
  studied in a wide area around  the supernova remnant W28. A 
  2\rlap{$^\circ$}.5
  $\times$ 2\rlap{$^\circ$}.5 field centered at $l$ = 6\rlap{$^\circ$}.5, $b$
  = 0$^\circ$ was surveyed using the Parkes 64-m radio telescope (HPBW
  14\rlap{$^\prime$}.7 at $\lambda$ 21 cm).  Even though W28
  is located in a complex zone of the Galactic plane, we have found
  different \HI\ features which are evidence of the interaction between W28
  and its surrounding gas. 
An extended cold cloud with about 70  M$_{\odot}$ of neutral hydrogen was
detected at the location of W28 as a self-absorption feature, near the
LSR velocity + 7 \ks. This \HI\ feature is the atomic counterpart of the
molecular cloud shown by previous studies to be associated with W28.
From this detection, we can independently confirm  a kinematical
distance of about 1.9 kpc for W28. In addition, the neutral hydrogen observed
in emission around the SNR displays a ring-like morphology in several
channel maps over the velocity interval [--25.0, +38.0] \ks. We propose
that these features are part of an interstellar HI  shell that has been
swept-up by the SN shock front. Emission from this shell is confused with
  unrelated gas. Hence, we derive an upper limit for the shell mass of 1200
  -- 1600  M$_{\odot}$, a maximum radius of the order of 20 pc, an
  expansion velocity of $\sim$ 30 \ks, an initial energy of about 1.4 --
  1.8 $\times 10^{50}$ ergs and an age of $\sim$ 3.3 $\times 10^4$ yrs. The
pre-existing ambient medium has a volume density of the order of
1.5 -- 2 cm$^{-3}$. W28 is probably in the radiative evolutionary phase,
  although it is not possible to identify the recombined thin neutral shell
  expected to form behind the shock front with the angular resolution of
  the present survey.  
\end{abstract}


\section{Introduction}
Each supernova remnant (SNR) is the unique product of its own history
(the progenitor and the explosion mechanism) and the characteristics of the
environs in which it evolves. The study of the interstellar medium around
SNRs can be used to understand  the appearance of a remnant in different
spectral regimes (distorted shapes, local brightness enhancements,
filamentary emission, etc.). Such studies 
also allow the analysis of the  temporal evolution of SNRs. In addition, 
the investigation of the gaseous matter around SNRs can lead to an
understanding of the Galactic interstellar medium. These studies are 
important in understanding the response of the interstellar gas to the large
injection of energy and momentum that a supernova (SN) explosion represents.

Numerous investigations of interaction of SNRs with the surrounding ISM
have been made using atomic and molecular lines (Routledge et al. 1991,
Pineault et al. 1993, Wallace et al. 1994, Frail et al. 1994, 1996 and
1998, Reynoso et al. 1995,  Dubner et al. 1998a and b). These
investigations show the manner in which the expansion of a SN
shock front modifies the surrounding environment and the effect that
the surrounding gas has, in turn,  on  the shape and dynamics of the 
SNR. In the present study, we report the results of an \HI\ study around 
the SNR W28 (G 6.4--0.1).

The SNR W28 is located in a very complex region of the Galaxy, near the
large HII regions M8 and M20, and the young clusters NGC 6530, NGC
6514, and Bo 14. It has a number of prominent morphological
characteristics. In the radio continuum, there is diffuse emission together
with thin filaments and small bright regions, as seen in Figure 1. This
image is the result of combining 50 VLA pointings into a 20 cm mosaic (Dubner
et al. 2000). In X-rays, diffuse thermal emission fills the
interior of W28, although ear-shaped segments of a limb-brightened shell
can also be observed toward the NE and NW (Rho et al. 1996). In the
optical, there are bright narrow filaments strongly
correlated with radio features and diffuse H$\alpha$ nebulosities,
apparently anti-correlated with the radio synchrotron emission (Long
et al. 1991, Dubner et al. 2000).

A number of observations support the existence of a physical
interaction between W28 and an adjacent molecular cloud: $(1)$ the
existence of shocked CO and other molecular species (Wootten 1981,
Frail \& Mitchell 1998, Arikawa et al. 1999) $(2)$ the detection of
over forty 1720 MHz OH masers distributed along the brightest
synchrotron features (Claussen et al.  1997 and 1999), 
and $(3)$ the coincidence of the molecular gas with the brightest
synchrotron filaments which, are the features with the flattest spectral
index in the SNR (as expected for high Mach number shocks; from Dubner et
al. 2000). All these indicators point to the
existence of an interaction between the SNR and the molecular cloud.

In what follows, we analyze the distribution of the neutral hydrogen
around W28, based on a survey of the $\lambda$21 cm \HI\ line carried
out for a 2\rlap{\arcdeg}.5 $\times$ 2\rlap{\arcdeg}.5 field with the
Parkes 64--m radio telescope.

\section{Observations and Data Reduction}

An area of 6.25 square degrees, centered at $l$ = 6\rlap{$^\circ$}.0, $b$
= $-$0\rlap{$^\circ$}.5 was observed using the Parkes 64m telescope on
June 23 and 24, 1995. The wide band (1.2--1.8 GHz) receiver was used,
with orthogonal, linearly polarized feeds. The half-power beam-width
of the telescope at the frequency of the \HI\ line is
14\rlap{$^\prime$}.7 and the pointing accuracy is $\leq$ 20
$^{\prime\prime}$. The system noise temperature is 28 K, measured
against cold sky. Each polarization was recorded with an instantaneous
bandwidth of 4 MHz over 2048 channels, giving a channel separation of
1.95 kHz (0.4 \ks\ at 1.4 GHz). The total velocity coverage is $\pm
400$ \ks centered at 0 \ks with respect to the Local Standard of Rest
(LSR).  After Hanning smoothing, the velocity resolution per channel
is 3.91 kHz (or 0.82 \ks).

In total, 289 positions were observed using constant Galactic latitude
scans with an integration time of 50 sec per spectrum. A reference
spectrum for band-pass calibration was taken using frequency switching
to $-$400 \ks at 25 minute intervals. Spectra were measured at the
Nyquist sampling interval, on a 7\rlap{$^\prime$}.5 grid.

Flux density calibration was made using scans across Hydra A, for
which the flux density was assumed to be 43.5 Jy at 1.4 GHz. The
brightness temperature scale was calibrated against the IAU standard
position S8, taken to have an integrated value of $897 \pm 66$ K\ks
(Williams 1973).  The conversion between flux density in Jy
beam$^{-1}$ and brightness temperature in K is 0.78 K/Jy beam$^{-1}$.
The rms uncertainty in the observations is 0.15 Jy, equivalent to 0.12
K. Initial processing was carried out using the SLAP and SDcube
software.  Final data analysis was performed using the AIPS software
package. All velocities used in this paper are referred to the LSR.

\section{The distance to W28 and the systemic velocity}

Previous distance estimates for W28 range between 1.3 and 3.6 kpc.
Milne (1970) estimated the distance to W28 to be 1.3--1.5 kpc. 
Lozinskaya (1974) derived a distance of 3.6 kpc based on H$\alpha$ 
measurements (assuming an LSR velocity near +18 \kms for W28).
Goudis (1976) estimated a distance of 1.8 kpc;
Clark \& Caswell (1976) suggested 2.3 kpc;  a further estimate by
Milne (1979) produced a distance of 2.4 kpc, and Venger et al. (1982) 
obtained a distance of
3 kpc based on \HI\ absorption measurements carried out with the RATAN-600
telescope.

On the other hand, OH (1720 MHz) maser emission associated with W28 was
detected by Frail, Goss \& Slysh (1994) at LSR velocities between +5
and +15 \ks, with most of the OH lines having velocities between +6 and
+8 \ks.  Strong OH absorption lines at 1612, 1665 and 1667 MHz were
reported at a radial velocity of +7.3 \ks (Goss 1968) while various
molecular species in the dense gas interacting with W28 have a central
velocity around +7 \ks (Pastchenko \& Slysh 1974, Wootten 1981,
Arikawa et al. 1999).  Arikawa et al. (1999) have shown the existence
of two different components in the CO emission at +7 \ks , a narrow
line corresponding to unshocked, quiescent molecular gas, and a broad
line, most likely arising from gas overtaken by the SNR shock. An
additional narrow CO component is detected at +21 \ks, but this line
probably originates in an unrelated cloud along the line of sight.
From these studies, we will adopt $\sim$ +7 \kms as the systemic
velocity of W28. For this LSR velocity, circular rotation models
provide near and far kinematic distances of 1.9 and 15 kpc. Because
independent estimates favor the lower value, we adopt a distance of
1.9$\pm$0.3 kpc for W28.

\section{The \HI\ around the SNR W28}

Figure 2 shows an \HI\ profile obtained after averaging spectra from the
entire observed region. The lower panel depicts the Galactic rotation
curve  toward $l$ = 6\rlap{\g}.5 $b$ = 0\g\ based on the model of Fich et
al. (1989), where $R_0=8.5$~kpc is assumed. 
The Galactic emission in this direction is mostly concentrated between
--50 and +50 \ks\ with a narrow absorption dip near +7 \ks. This
is a very strong self-absorption feature produced by an unusually cold
cloud that extends over a large region (covering at least 20$^\circ$ of
longitude in the direction of the Galactic center), and including the
direction of W28 (Riegel \& Jennings, 1969).

Figures 3 and 4 show the distribution (in greyscale and 
white contours) of the \HI\ emission within the velocity interval where
significant \HI\ emission is observed. In order to  compare radio continuum
and \HI\ structures, the black contours represent the boundaries of the
radio continuum emission associated with W28 (smoothed to the resolution of
the \HI\ data). The other bright continuum sources plotted in the field are
the Trifid Nebula (M20; G07.00--0.3) to the left of W28, and the compact
HII region W28 A--2 (G05.89--0.4) to the lower right corner of the
Figures. The average \HI\ field emission has been subtracted from all the
images for presentation purposes. The greyscale plotted along the upper
edge of Figures 3 and 4 is kept constant in all images in order to emphasize 
changes in structures for different velocities. With the exception of the
first image of Figure 3 and the two last images of Figure 4,  where the
integration intervals were chosen to be 35 and 30 \ks, respectively, the
remaining images result from the average over 5 \ks (6 consecutive spectral
channels). The central velocity of each  integration interval is
indicated in the top right corner of each panel. 

The analysis of the \HI\ distribution around W28 is quite complex, because
of its location close to both the Galactic center and the Galactic plane.
To identify structures that may be associated with the SNR, we look for
features that may reveal the impact of the SNR expansion on the
surrounding interstellar medium (ring-shaped \HI\ structures,
expanding \HI\ caps, etc.). Also, we have attempted to locate \HI\ 
concentrations that appear to be associated with prominent features
observed in the radio continuum emission of W28.
 
Bright \HI\ emission is observed at negative velocities. Based on Galactic
circular rotation models, negative velocities should arise from gas at 
distances $>$ 17 kpc.  However, it is unlikely that the large bright structures
observed between \vlsr $\sim$ --30 and --2 \ks\ correspond entirely to this
distant gas. We have analyzed the HI features in this velocity range in an
attempt to find associations with W28. The negative velocities could
result from kinematic perturbations arising from the SNR.

Between $\sim$ $-$32 and $\sim$ $-$2 \ks (Fig. 3), the brightest \HI\ 
emission regions are preferentially distributed around W28, encircling
the source along different sides.   Particularly, 
at \vlsr= $-$27.5 \kms and \vlsr=
$-$22.5 \kms HI concentrations can be observed, forming a clumpy
incomplete shell. It is possible that part of this gas had been swept-up
by the expanding SN shock, although a ``cap''--like feature would be 
the expected morphology for associated \HI\ concentrations at these
high negative velocities. Thus the association is uncertain. 
At \vlsr= $-$7.5 and $-$2.5 \kms a good
coincidence is observed between \HI\ concentrations and some distortions
observed in
the outer envelope of W28 along the E and N sides. These structures
are compatible with the hypothesis that the expanding shock wave of
the SNR is pushing the interstellar neutral gas outwards.

At \vlsr= +2.5 \kms, the \HI\ surrounds W28 with an almost complete
shell-like structure. A striking morphological correspondence is
observed between the two \HI\ maxima to the N and NE of W28, at ($6\arcdeg
50\arcmin, + 0\arcdeg15\arcmin$) and (6$\arcdeg 50\arcmin,
-0\arcdeg20\arcmin$) respectively, and the two
sites where the radio continuum shell is indented.  The \HI\ 
concentration near (6\arcdeg50\arcmin,--0\arcdeg20\arcmin) coincides
with the CO concentration reported by Arikawa et al.  (1999) to be
quiescent molecular gas associated with W28.  This \HI\ component is
also present in the following channel image at \vlsr= +7.5 \kms (top
left image of Figure 4). However, at \vlsr= +7.5 \kms the HI feature in
emission is masked by the strong absorption toward the SNR.

As mentioned before, in the image centered at \vlsr= +7.5 \ks (Figure 4) 
the most remarkable
feature is the central depression observed in \HI\ emission. This \HI\ 
depression results from self--absorption produced by an extended, cold
cloud centered near +7 \kms, which is part of the complex of cold
clouds reported by Riegel \& Jennings (1969) in this direction of the
Galaxy.  To analyze the absorption features, in Figure
5 we display the negative contours (white lines) overlapping W 28
(greyscale)  as obtained from an integration of \HI\ between +4 and +9 \kms. 
Based on Figure 5 we can conclude: (1) that the cold cloud is much larger
than the SNR, and (2) the location of the deepest absorption hole does
not coincide with the brightest synchrotron feature to the E of W28,
but it approximately overlaps the thin radio
filament that crosses W28 in the E-W direction (see Figure 1). This
synchrotron filament has been shown by Dubner et al. (2000) to have
the flattest spectral index of all parts of the SNR. Associated with
this filament, Arikawa et al. (1999) have shown the existence of
shocked CO gas (with broad wings, between \vlsr~ -40 and +40 \kms) and 
unshocked CO gas
(between \vlsr~ +4 and +9 \kms). The shocked CO is associated with 
numerous OH (1720 MHz) masers (Claussen et al. 1997 and 1999). For the
unshocked gas, Arikawa et al. (1999) estimate a kinetic temperature $\leq$
20 K, a density $\leq 10^3$ cm$^{-3}$ and a total H$_2$ mass of 4000
M$_\odot$. For the shocked gas the physical parameters are: kinetic
temperature $\geq$ 20 K,  density $\geq 10^4$ cm$^{-3}$ and a total H$_2$
mass of 2000 M$_\odot$. We conclude that we have detected
the atomic hydrogen counterpart of the unshocked molecular cloud associated
with W28, thus confirming on the basis of \HI\ data the systemic velocity 
and the distance of 1.9 $\pm$ 0.3 kpc for W28. From the present data 
we estimate that the  mass of the cold \HI\ responsible for the
self-absorption  is 70 M$_\odot$. Therefore, only a small fraction of the
total molecular gas mass is detected in atomic form.

At \vlsr = +17.5 \kms a conspicuous  \HI\ shell centered near ($l\sim 6\arcdeg
40\arcmin , b\sim 0\arcdeg 12\arcmin$, with radius $\sim$ 0\arcdeg.6) 
is observed surrounding the SNR.  The presence of these concentrations
distributed in a ring-like shape, together with the other features
previously described at negative LSR velocities,  suggest that part of the
surrounding \HI\ gas may have been swept-up by the expanding SNR shock
wave, forming a thick \HI\ interstellar shell. The existence of diffuse
thermal X-ray emission filling the interior of W28 (Rho et al. 1996),
supports the hypothesis that the center has been evacuated. 

At higher positive velocities the
most noticeable feature which may be associated with W28 is the central
concentration present at \vlsr= +32.5 and +37.5 \kms. This
concentration appears to be projected onto the interior of the SNR and
can be interpreted as the ``cap'' of the \HI\ shell 
expanding around W28.
 
Based on these results we can propose a  model where the explosion took 
place near ({\it l, b},V) = (6\arcdeg
30\arcmin , $-$0\arcdeg 12\arcmin, +7 \kms).  The present shell radius
is $\sim$ 0\arcdeg.6, or 20 pc at a distance of 1.9~kpc. On the basis of
the adopted systemic velocity of +7 \kms and the presence of a
``cap''-like feature near +37 \kms, the expansion
velocity of this structure is estimated to be about 30$\pm$3 \kms.

The total associated \HI\ mass  can be estimated by integrating the
contributions of all the structures which are considered to be part of the
atomic gas shell. In other words, the \HI\ emission between approximately
V$_{\rm LSR} \sim -25$ \kms and V$_{\rm LSR} \sim +37$ \kms,
which appear encircling W28 in the different channel maps, plus the
features projected onto the center of the SNR at the high positive
velocities. After the subtraction of an appropriate background contribution
(assumed to be 3--$\sigma$ below the isocontour which outlines the
associated features), an \HI\ mass of  
1600$\pm$240 M$_\odot$ is obtained. The quoted error takes into account
the uncertainties in the selection of the boundaries for the integration.
This is an upper limit for the associated mass, since it is
impossible to separate the contributions from unrelated \HI{}.  In this
total mass estimate, about $\sim 400$ M$_\odot$ correspond to the 
features  at velocities more negative than --7.5 \kms, whose
association with W28 may be questionable. Thus the total swept-up \HI\ mass
can vary between $\sim$ 1200 and 1600 M$_\odot$.

Assuming that 
the mass is uniformly distributed over a sphere of radius of 20 pc,
we obtain an upper limit for the ambient interstellar medium (ISM)
density of $\sim$  1.5 --  2.0 cm$^{-3}$ (depending on the value used for
the total mass).  The kinetic energy would be E$_{\rm kin} \leq$ 1 $\times$
10$^{49}$~ergs and the initial energy of the explosion about 1.4 -- 1.8
$\times$ 10$^{50}$~ergs (based on Chevalier's  1974 model). Given these 
values for the 
initial explosion
energy and the unperturbed ISM density, and by considering
the following expression (Rohlfs \& Wilson 1996):

\begin{equation}
\tau_{rad} = \biggl[ {{4.56\times 10^7}\over{(1+x_H)\ T_4}}
\biggl({E_{51}\over n_o}\biggr)^{2/5}\biggr]^{5/6}\ yr,
\label{trad}
\end{equation}

\noindent we can estimate the time for the onset of the radiative phase of
SNR evolution to be $\tau_{rad}$  $\sim 2.1 -2.4 \times 10^4$~yr, while   
$R_{rad} \sim 10$~pc is the radius of the SNR at that stage.
In Eq.(1) $T_4$ is the temperature just behind the SNR shock wave (in 
units of 10$^4$~K) which was set to 100 ( radiative losses start to be 
an important process at about 10$^6$~K), $E_{51}$
is the initial SN energy in units of 10$^{51}$~erg, $x_H$ is the
ionization fraction (assumed to be 0) and $n_0$ is the 
unperturbed ISM density in \cmm .

By assuming a radius of about 13 pc for W28 (from an angular size of $\sim$
23.5 arcmin and a distance of $\sim$ 1.9 kpc), we can conclude that this
remnant is well in the radiative phase of evolution and has a current age
of $3.3 \times 10^4$~yr, which is in good agreement with previous age
estimates (Frail et al.~1993, Vel\'azquez 1999). In this evolutionary
stage, it is expected that a thin \HI\ shell forms by recombination behind
the shock front with a width of about 10\% of the radius.

\section{Conclusions}

We have carried out a study of the neutral hydrogen in the environs of the SNR
W28.  Our analysis of the kinematics and distribution of the \HI\  has
revealed several \HI\ features that are most probably with W28, revealing
signatures of the interaction of this SNR with the interstellar medium. We
have detected, as a self-absorption feature around $\sim$ 7 \kms,  the
neutral gas counterpart of the molecular cloud detected by Arikawa et al.
(1999) in unshocked CO gas. Based on the presence of this cold cloud, we
can independently confirm a kinematical distance of 1.9$\pm$0.3 kpc for W28.

Portions of an incomplete \HI\ shell are also observed in emission at
different positive and negative LSR velocities, with a maximum
angular size ($\sim 0^\circ.6$) at V$_{LSR}$=+17.5\kms.  
An \HI\ cloud is detected near V$_{\rm LSR} \sim +37$ \kms
overlapping the center of W28. We interpret this last feature as the
``cap'' of the irregular expanding interstellar shell swept-up by the W28
shock front. The mass of this shell has been estimated to be between
1200 and  1600 M$_{\odot}$.  

Based on the present results, the following scenario for W28 can be
proposed:

(a) A SN explosion of energy $\sim$ 1.6 $\times 10^{50}$~ergs occurred
about 3.3 $\times 10^4$~yr ago, at the position ({\it l, b}) = 
(6\arcdeg30\arcmin
, $-$0\arcdeg 12\arcmin) and at distance of $\sim$~1.9 kpc. At this location,
the ambient density of the ISM was $\sim$1.5 -- 2 cm$^{-3}$.

(b) The expanding shock wave has collided with a cold gas
concentration, observed as an absorption \HI\ feature and as molecular clouds
around  the LSR velocity of 7 \kms.   The mass of this cold cloud
($\sim $ 70 M$_\odot$), is only
a small fraction of the total mass estimated for  molecular
hydrogen. Most of the atomic hydrogen is detected in emission as an 
\HI\ shell, as mentioned below.

(c) The interaction of the SN shock front with the surrounding \HI\ gas
has swept-up a thick interstellar HI shell, presently expanding at $\sim$ 
30 \kms. W28 has entered into the radiative stage of evolution about
2$ \times 10^4$ yrs ago. However, the thin neutral shell expected to form by
recombination behind the shock front could not be identified
because of the relatively low angular resolution of the present study.

\acknowledgments 

We thank the anonymous referee for helpful and constructive comments. PFV
acknowledges the partial financial support from CONACyT (M\'exico) grant
36572-E. The research presented in this work was also funded by a
Cooperative Science Program between NSF and CONICET (Argentina) and
PIP-CONICET grant 4203/96. 
The Very Large Array of the National Radio Astronomy Observatory is a
facility of the National Science Foundation operated under cooperative
agreement by Associated Universities, Inc.
The Australia Telescope is funded by the Commonwealth of Australia for 
operation as a National Facility, managed by CSIRO.

\clearpage



\section{Figure captions}
\figcaption[fig1]{Radio continuum image of the SNR W28 and the HII
regions M20 and W28A--2,  and the SNR G7.06--0.12. The greyscale range is
[0, 0.5] Jy beam$^{-1}$. This image was obtained with the VLA at 1415 MHZ
by Dubner et al. (2000). The resolution is  88\arcsec $\times$ 48\arcsec,
P.A.=8\arcdeg. The $1\sigma$ noise level is 5 mJy beam$^{-1}$.}


\figcaption[fig2]{{\bf (a)} Average HI emission profile at 
$l \simeq$ 6\fdg 5 and $b\simeq$ 0\arcdeg\ from the current data; {\bf (b)}
Galactic rotation curve at $l=6\fdg 5$, $b=-0\fdg 1$ (Fich et al. 1989).}


\figcaption[fig3]{Images of \HI\ emission (in grey and white
contours) between $-$55 and +3.\ks. The first image is an
 integration from $-$70 to $-$35 \ks. In the subsequent images the
 integration is carried out over an interval of 5 \ks. The greyscale range 
is [--5, 15] Jy beam$^{-1}$ \ks , while the white contours 
correspond to  --2, 2, 6, 10, 14, 18, 22 and 26 Jy beam$^{-1}$ \ks .
The black contours show the 2, 2.5 and 3 Jy beam$^{-1}$ levels  of the 
radio continuum of W28 at 1415 MHz (from Dubner et al. 2000), with an angular
 resolution of 14\rlap{\arcmin}.7 . Each panel is labeled with the central
 velocity of th eparticular interval of integration. The arrows on the left
 top corner of the first panel show the N and E directions in the J2000
 Equatorial Coordinate system,  to facilitate comparison with Figure~1.}


\figcaption[fig4]{As in Fig.~3, \HI\ images in the velocity range
[+7.5, +85] \ks . The last two images are obtained from integration over
an interval of 30 \ks .}


\figcaption[fig5]{Overlay of the \HI\ depression integrated between
4 and 9 \ks\ (white contours), and the W28 radio continuum (greyscale).
The contours correspond to the $-$26, $-$22, $-$18, $-$14, $-$10, $-$6, 
$-$2 and 2 Jy beam$^{-1}$ \ks\ levels.
The radio continuum of W28 (with a resolution of 88\arcsec $\times$ 
48\arcsec ) is shown with a greyscale range of [$-$0.06, 0.75] Jy beam$^{-1}$.}

\end{document}